# All-optical discrete illumination-based compressed ultrafast photography

Long Cheng[1], Dalong Qi[1,*], Jiali Yao[2,5], Ning Xu[1], Chengyu Zhou[1], Wenzhang Lin[1], Yu He[1], Zhen Pan[1], Hongmei Ma[1], Yunhua Yao[1], Lianzhong Deng[1], Yuecheng Shen[1], Zhenrong Sun[1], Shian Zhang[1,3,4,*]

[1]*State Key Laboratory of Precision Spectroscopy, School of Physics and Electronic Science, East China Normal University, Shanghai 200241, China*

[2]*College of Science, Shanghai Institute of Technology, Shanghai 201418, China*

[3]*Joint Research Center of Light Manipulation Science and Photonic Integrated Chip of East China Normal University and Shandong Normal University, East China Normal University, Shanghai 200241, China*

[4]*Collaborative Innovation Center of Extreme Optics, Shanxi University, Taiyuan 030006, China*

[5]*e-mail: jlyao@sit.edu.cn*

*Corresponding authors: dlqi@lps.ecnu.edu.cn, sazhang@phy.ecnu.edu.cn*

## Abstract

Snapshot ultrafast optical imaging (SUOI) plays a vital role in capturing complex transient events in real time, with significant implications for both fundamental science and practical applications. As an outstanding talent in SUOI, compressed ultrafast photography (CUP) has demonstrated remarkable frame rate reaching trillions of frames per second and sequence depth over hundreds of frames. Nevertheless, as CUP relies on streak cameras, the system's imaging fidelity suffers from an inevitable limitation induced by the charge coupling artifacts in a streak camera. Moreover, although advanced image reconstruction algorithms have improved the recovered scenes, its high compression ratio still causes a compromise in image quality. To address these challenges, we propose a novel approach termed all-optical discrete illumination compressed ultrafast photography (AOD-CUP), which employs a free-space angular-chirp-enhanced delay (FACED) technique to temporally stretch femtosecond pulses and achieves discrete illumination for dynamic scenes. With its distinctive system architecture, AOD-CUP features adjustable frame numbers and flexible inter-frame intervals ranging from picoseconds to nanoseconds, thereby achieving high-fidelity ultrafast imaging in a snapshot. Experimental results demonstrate the system's superior dynamic spatial resolution and its capability to visualize ultrafast phenomena with complex spatial details, such as stress wave propagation in LiF crystals and air plasma channel formation. These results highlight the potential of AOD-CUP for high-fidelity, real-time ultrafast imaging, which provides an unprecedented tool for advancing the frontiers of ultrafast science.

1.Introduciton

Ultrafast optical imaging (UOI) of transient events at their native timescales is of great scientific significance and practical value. Over the past decade, single-shot ultrafast imaging with multiple frames[1] has emerged as a pivotal direction in the field of UOI. Aimed at capturing the spatiotemporal evolution of non-repeatable or hard-to-reproduce ultrafast phenomena within a single exposure, this approach overcomes the inherent limitations of conventional pump-probe techniques[2] that rely on repetitive measurements. It has become a key tool for real-time observation and analysis of complex transient dynamics, such as rogue wave generation[3], ultrafast light field evolution[4], ultrafast laser-matter interaction[5].

In recent years, a variety of single-shot UOI techniques have been proposed, which can be broadly categorized into two categories. One is framing photography[6], which separate different temporal frames and project them onto distinct detectors or regions of a detector array in space or spatial frequency, such as sequentially timed all optical mapping photography (STAMP)[7], framing imaging based on noncollinear optical parametric amplification (FINCOPA)[8], and femtosecond videography for atomic and molecular dynamics (FRAME)[9]. The other includes techniques relying on compressed sensing (CS), which utilize optical encoding and computational reconstruction to retrieve dynamic scenes, with notable examples including compressed ultrafast photography (CUP)[10, 11], compressed ultrafast spectral-temporal photography (CUST)[12] and swept coded aperture real-time femtophotography (SCARF)[13]. Among these, CUP, as a representative computational ultrafast imaging technique, achieves imaging speeds up to trillions of frames per second (fps) within a single measurement. CUP integrates spatial encoding with streak imaging[14] to passively compress a three-dimensional temporal scene into a two-dimensional (2D) image, which is subsequently reconstructed computationally to reveal the full temporal evolution of the scene. Based on this imaging modality, CUST utilizes temporally chirped-pulse illumination to map time to wavelength, thereby transforming temporal information into spectral information, and further reconstruct ultrafast dynamics at a frame rate up to $3.84 \times 10^{12}$ fps using CS-based algorithms. SCARF, achieves full pixel-wise encoded depth in a single-shot ultrafast imaging acquisition by employing a single chirped pulse and a pulse shaping module. By leveraging temporal-spectral mapping and spectral-spatial scanning, SCARF links the full pixel-wise encoded aperture to a conventional CCD camera, enabling a real-time imaging speed up to $156.3 \times 10^{12}$ fps. Compared with ultrafast framing imaging methods, techniques such as CUP employ spatial multiplexing for temporal frames during data acquisition, offering significant advantages in data throughput and space-time bandwidth product[15].

In recent advances, CUP and its variants have demonstrated remarkable versatility across multiple imaging scales[16-20], from macroscopic to microscopic regimes, including unprecedented scenes such as propagation of optical Mach cones[21], laser-induced ablation dynamics[22], spatiotemporal evolution of complex ultrafast light fields[23], and snapshot fluorescence lifetime imaging microscopy[24]. However, the inherently high compression ratio in CUP limits the fidelity of image reconstruction. To address this issue, researchers have sought to optimize performance through algorithmic improvements[25, 26] and novel optical encoding strategies[27, 28]. As for algorithm improvement, Jin *et al*. combines the strengths of training-free self-supervised neural networks and traditional iterative algorithms by integrating the plug-and-play (PnP) framework with deep image prior (DIP), leading to a novel hybrid reconstruction method termed PnP-DIP, which enables fast and efficient feature extraction directly from raw measurement data, yielding high-quality image reconstructions. As for hardware implementation, Yao *et al.* introduced discrete illuminated-based compressed ultrafast photography (DI-CUP)[29], which utilizes a train of sub-pulses with adjustable number and temporal spacing to discretely illuminate dynamic scenes. DI-CUP effectively distributes the temporal evolution of the scene across multiple sub-pulses, thereby flexibly adjusting the compression ratio, suppressing inter-frame crosstalk, and enhancing the fidelity and robustness of image reconstruction. Despite its conceptual and performance breakthroughs, DI-CUP remains constrained by the inherent limitations of streak camera-based systems, including high cost and charge coupling effects, which inevitably degenerate both image quality and spatial resolution.

To address the above-mentioned limitations in CUP, we propose an all-optical discrete illumination-based CUP (AOD-CUP). In this method, temporal chirping and ultrafast pulse train generation are integrated for discrete illumination. Specifically, a free-space angular-chirp-enhanced delay (FACED)[30] device is utilized to stretch a single femtosecond pulse and reshape it into a pulse train in the time domain for discrete illumination, which significantly reduces the compression ratio and suppressed inter-frame overlap. In addition, the different sub-pulses in the pulse train generated by FACED have different wavelengths, which allows dispersive elements to replace the streak camera to realize time-shearing to avoid image aberrations caused by the space-charge effect, thus further ensuring high-fidelity ultrafast imaging. A dynamic imaging quality comparison of AOD-CUP against that of DI-CUP using test targets was implemented to demonstrate the superior spatial fidelity of our approach. Furthermore, we applied AOD-CUP to visualize the transient stress wave propagation[31] within a LiF crystal induced by a femtosecond laser, as well as the formation of an air plasma channel, with result consistent with conventional pump–probe measurements. These demonstrations underscore the significant advantages and broad application potential of AOD-CUP for probing complex ultrafast phenomena.

## 2. Principles and Methods

### a. Experimental configuration

The experimental setup of AOD-CUP is illustrated in Fig. 1. A Ti:sapphire femtosecond laser (Coherent, Legend Elite USP-HE) with a central wavelength of 800 nm, spectral bandwidth of about 40 nm, and pulse duration of 50 fs is employed as the illumination light source. Initially, a femtosecond pulse is directed into FACED system through a beam splitter (BS1), which consists of a diffraction grating (G1, LBTEK, BG25-1200-750) and a pair of long highly reflective plane mirrors separated by a tunable distance. Upon passing through the grating, the spectrum of the femtosecond pulse is spatially dispersed and subsequently different wavelength components are reflected between the two mirrors at different angles of incidence. Due to a slight angular misalignment between the mirrors, some specific wavelength components undergo multiple reflections resulting in the incident angle being perpendicular to a mirror and eventually return to BS1 along the original paths. Among them, different wavelength components undergo different numbers of reflections to achieve perpendicular incidence, corresponding to different optical path differences, and thus a pulse train is generated. The FACED system enables the generation of pulse trains with a tunable number of sub-pulses and inter-pulse intervals ranging from tens of picoseconds to nanoseconds. The number of sub-pulses $M$ is simply proportional to the input cone angle of the beam $\Delta\theta$ [32]:

$$M = \frac{\Delta\theta}{\alpha}, \tag{1}$$

where $\alpha$ is the small angular misalignment between the two long plane mirrors. The temporal interval between adjacent sub-pulses can be expressed as:

$$\tau = \frac{2S}{c}, \tag{2}$$

where $S$ denotes the distance between the two long plane mirrors and $c$ is the speed of light. The generated pulse train is then directed back to BS1, reflected by a mirror (M, Thorlabs, BB1-E03), and used to illuminate a dynamic scene. The pulse train carrying the scene is collected by a camera lens and subsequently split by BS2 into two paths. One path is directed to an external camera (Basler, acA1920-40gm) for spatiotemporal integration image acquisition, while the other is projected onto a mask with a pseudo-random binary pattern for encoding. The encoded scene is further relayed by a lens (L1, LBTEK, AD512-B, $f$ =75 mm) to a second diffraction grating (G2, Thorlabs, GTI25-03) for spectral shearing, and finally imaged onto a camera (EMCCD, Andor iXon+ 888) through another relay lens (L2, LBTEK, AD512-B, f =100 mm), enabling single-shot acquisition of the ultrafast scene.

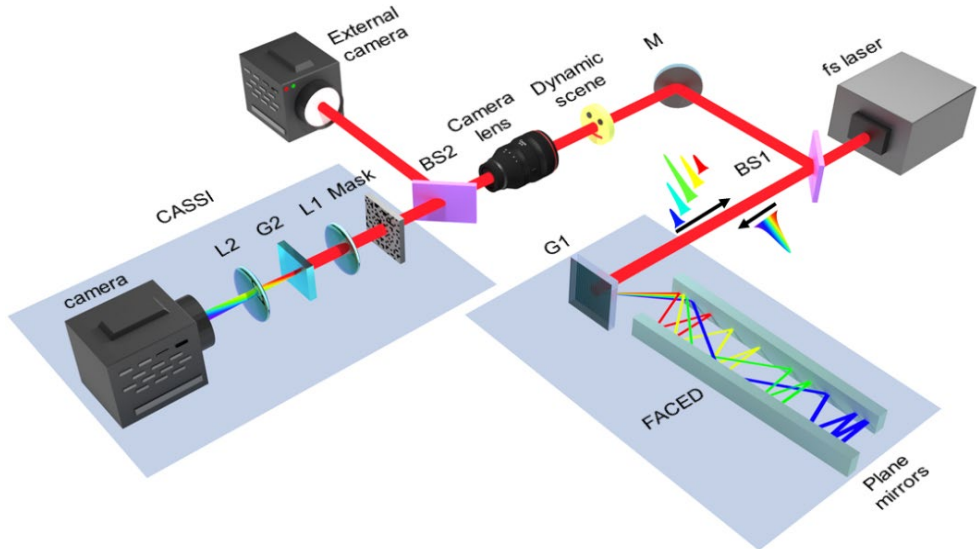

**Fig. 1.** Experimental configuration of AOD-CUP. BS1, BS2, beam splitters; L1, L2, lenses; M,

mirror; G1, G2, gratings.

**b. Imaging model**

The forward data acquisition of AOD-CUP is similar to that of DI-CUP, as illustrated in Fig. 2(a). An ultrafast pulse sequence represented by an operator **D**, matching to the temporal scale of the dynamic scene, is employed to illuminate the dynamic scene $I(x, y, t)$. This illumination is then encoded by a pseudorandom binary encoding mask **C**, followed by a spatial shearing operation **S** introduced by a diffraction grating. An exposure of the camera compresses the dynamic scene through a spatiotemporal integration **T** into a 2D measurement $E(x', y')$. Therefore, the forward data acquisition process can be mathematically described as:

$$E(x', y') = \mathbf{TSCD}I(x, y, t). \quad (3)$$

For image reconstruction, recovering the original dynamic scene from the compressed 2D image constitutes an ill-posed optimization problem. To find the optimal solution, CS algorithms are typically employed to minimize the objective function $f(I)$, which can be formulated as:

$$I^* = \arg\min_{I} f(I) = \arg\min_{I} \frac{1}{2} \|E - \mathbf{TSCD}I\|_2^2 + \lambda \Phi(I), \quad (4)$$

where $\|E - \mathbf{TSCD}I\|_2^2$ represents the fidelity term, $\Phi(I)$ is the regularization term, and $\lambda$ is the regularization parameter, which effectively balances the two terms and optimizes the quality of the result. In this work, we address this challenge by solving the objective function using the total variation and cascaded denoisers-based (TV-CD) algorithm in a plug-and-play [33] framework, as shown in Fig. 2(b). The TV-CD algorithm adopts a cascaded denoising configuration, consisting of TV and the fast and flexible denoising convolutional neural network (FFDNet)[34], the dilated-residual U-Net denoising neural network (DRUNet)[35] and the fast deep video denoising network (FastDVDNet)[36]. Here, FFDNet and DRUNet achieve effective denoising of single frame in dynamic scenes by take the noise level maps as auxiliary inputs and FastDVDNet further leverages temporal neighborhood information to enhance the temporal consistency of residual noise in the output frames, thereby significantly improving overall denoising performance in dynamic scenarios. In this architecture, each denoiser is assigned an independent number of iterations. Once a denoiser completes its designated iterations, the system proceeds to the next denoiser in the sequence, continuing this process until all iterations are completed. During the entire iterative reconstruction, the noise standard deviation $\sigma$ is adaptively updated based on the residuals, thereby progressively enhancing the quality of the reconstructed images.

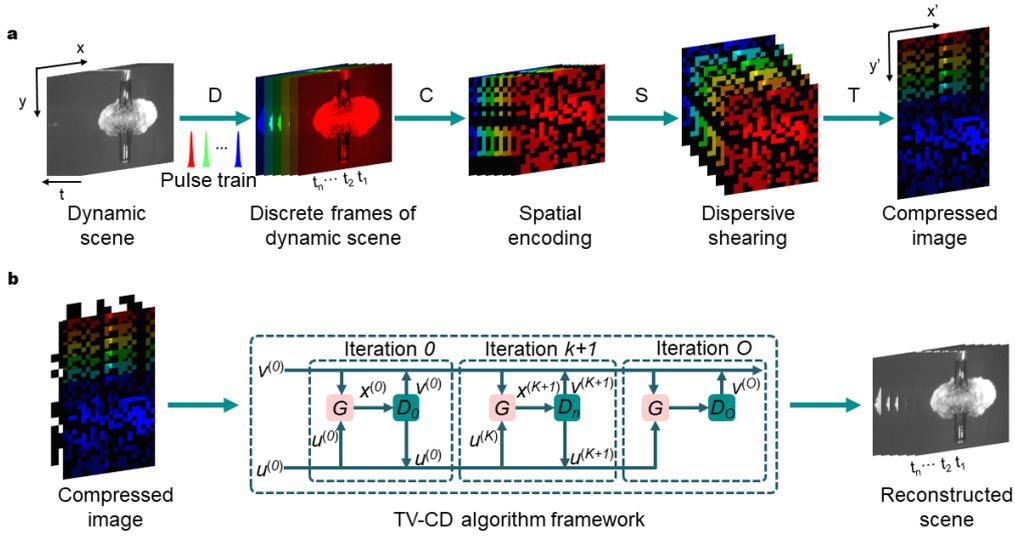

**Fig. 2.** Flowchart of the data acquisition and image reconstruction. (a) Schematic diagram of the data acquisition for AOD-CUP, where $x$, $y$: spatial coordinates of the dynamical scene; $t$: time; $x'$, $y'$: spatial coordinates at the camera. (b) Data flowchart of the TV-CD algorithm for AOD-CUP reconstruction.

## 3. Results and Discussion

### a. Characterization of the reconstructed image quality in AOD-CUP

First, we demonstrated the superior dynamic spatial resolution of AOD-CUP by comparing it against DI-CUP. A USAF 1951 resolution test target was selected as an object for ultrafast illumination. In the AOD-CUP system, the number of sub-pulses was set to 12 by tuning $S$ and $\alpha$ in the FACED module, yielding an inter-pulse interval of approximately 80 ps and a total duration of 960 ps. To ensure an equitable comparison with DI-CUP, a 10× objective lens (Olympus, PLN 10×) was used for microscopic imaging (resulting in a system magnification of ~12.8×, which is consistent with that of DI-CUP) and a pseudo-random spatial encoding mask with 10 μm unit was applied to the resolution target. Being spectrally dispersed by the grating, the pulse train was spatiotemporally captured by the camera. Subsequently, the original scene was reconstructed using the TV-CD algorithm, and Fig. 3(a) presents a representative reconstructed frame that containing two sets of horizontal and vertical bar patterns, corresponding to elements 7-1 and 7-2 of the resolution target. The results indicate that in element 7-1, both the horizontal and vertical line structures are clearly resolved, whereas the lines in element 7-2 are not distinguishable. To quantitatively evaluate the spatial resolution of the reconstructed image, the intensity profiles of the horizontal and vertical features are respectively integrated along their dashed lines, and their normalized profiles are shown in Fig. 3(b). This suggests that the spatial resolution of AOD-CUP reaches approximately 128 lp/mm, which significantly outperforms the resolution limit of 45.3 lp/mm in DI-CUP. Furthermore, a single-frequency Ronchi ruling was employed as the object to evaluate the system's spatial distortion. One of the reconstructed images is shown in Fig. 3(c) as the representative. For a more intuitive analysis, 2D Fourier transforms (FTs) were performed on both the reconstructed image and the spatiotemporal integrated image as the ground truth, with their corresponding results shown on the left and right sides in Fig. 3(d), respectively. In addition, the intensities of the FT images along the $y^{-1}$-axis direction were extracted. As can be observed in Fig. 3(e), the spatial frequency distribution of the AOD-CUP image closely matches that of the static reference, indicating minimal distortion. Therefore, the reconstructed image quality experiment confirms that AOD-CUP achieves a ~1.82× improvement in spatial resolution compared to DI-CUP, as well as a negligible spatial distortion, validating its outstanding capability in delivering high-fidelity, ultrafast dynamic imaging.

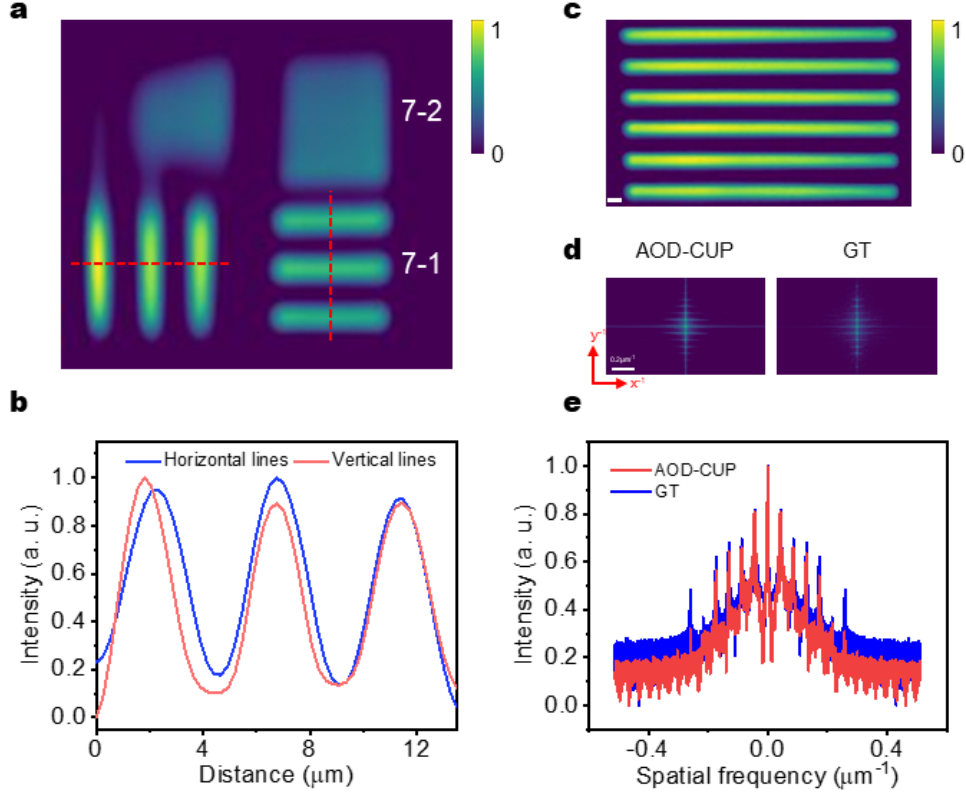

**Fig. 3.** Characterization of the reconstructed image quality in AOD-CUP. (a) One representative reconstruction frame of the resolution target; (b) Intensity distribution curves along the horizontal and vertical lines on the 7-1 element in (a); (c) One representative reconstruction frame of the single-frequency Ronchi ruling; (d) 2D Fourier transforms (FTs) of the reconstructed image (left) and the spatiotemporal integrated image (right); (e) The integrated intensity of the FT images along $y^{-1}$-axis. Scale bar: 10 μm.

### b. Comparative measurements of spatiotemporal chirped light fields

Spatiotemporally chirped light fields play a pivotal role in various domains such as ultrafast laser processing[37] and manipulate spatiotemporal foci[38]. To investigate the wavefront evolution of such fields, we experimentally constructed a dynamic scene by combining a discrete pulse train generated by FACED with a dual-grating configuration, as illustrated in Fig. 4(a). Each sub-pulse produced by the FACED module was directed through a pair of diffraction gratings, whose groove densities were 1200 gr/mm and 600 gr/mm for comparison. Specifically, the introduced spatial chirps cause different wavelengths to be transversely dispersed in space, enabling the sub-pulses to sequentially scan across a pattern of letters before being captured by the subsequent hyperspectral imaging system. Owing to the fact that sub-pulses of different wavelengths illuminate the pattern at distinct spatial locations and temporal instants, the spatial profile of the output field evolves over time, thereby capturing the intrinsic spatiotemporal correlation of each light field in a snapshot. 5 over all the reconstructed images for the 1200 gr/mm and 600 gr/mm cases are shown in Figs. 4(b) and 4(c), respectively. As the illumination wavelength shifts from long to short, the sub-pulses sweep across the pattern from left to right in a sequential manner. Additionally, we calculated the temporal evolutions of the beam centroids along the *x*-direction for the two spatiotemporally chirped beams, as shown in Fig. 4(d). Due to the higher spatial chirp introduced by the 1200 gr/mm grating pair compared to the 600 gr/mm one, the corresponding scanning speed of the former is significantly increased from 0.36 mm/ns to 0.63 mm/ns.

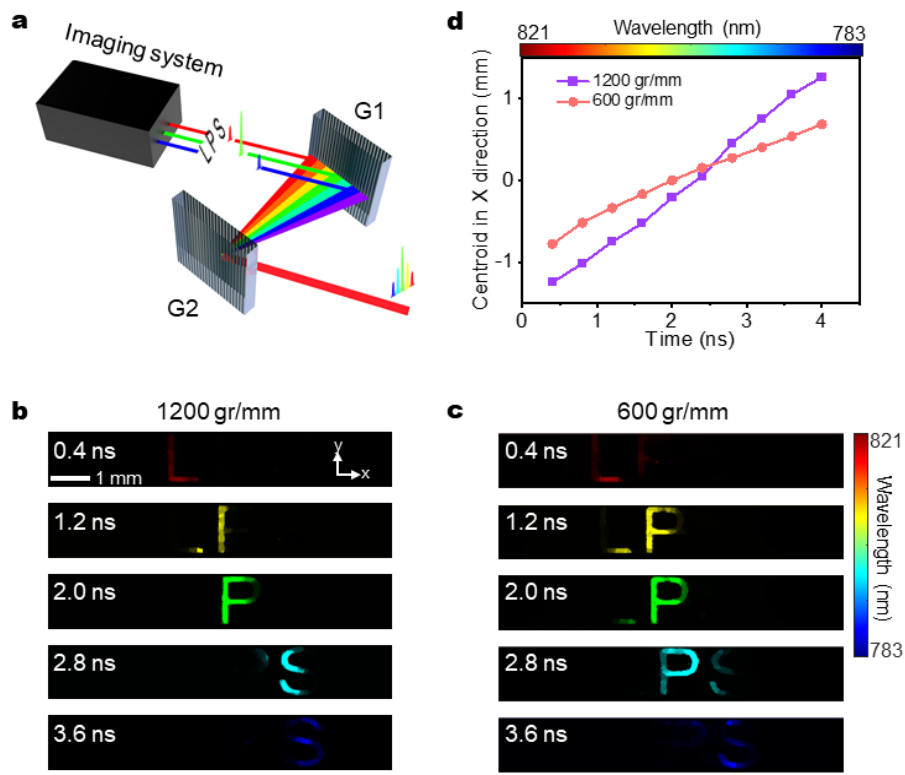

**Fig. 4.** AOD-CUP imaging of spatiotemporally chirped light fields. (a) Experimental configuration of the dynamic scene. (b) Selected frames from the reconstruction, showing a spatiotemporally chirped pulse train sweeping across a set of letters, with the grating pair used being 1200 gr/mm; (c) Selected frames from the 10-frame reconstruction, showing a spatiotemporally chirped pulse train sweeping across a set of letters, with the grating pair used being 600 gr/mm; (d) Temporal evolution of the calculated *x*-direction centroid of the illumination beam; Scale bar: 1 mm.

### c. Observation of laser-induced stress waves

Femtosecond laser processing of transparent solid materials[31] has been extensively studied. When an intense ultrafast laser pulse interacts with a single-crystal sample, it generates stress waves in the photoexcited region. Observing the transient stress distribution immediately after photoexcitation is critical for understanding the mechanisms of laser-induced material deformation and for developing high-precision material processing techniques. Here, we employed an AOD-CUP system to visualize the transient stress wave propagations induced by femtosecond laser excitation inside different types of LiF crystals. The experimental configuration is illustrated in Fig. 5(a). An ultrafast pulse delivered from the Ti:sapphire laser is split into two paths. The pulse in one path is frequency-doubled using a $\beta$-BBO crystal to produce a 400 nm femtosecond pulse with an energy of 600 μJ for pump. Subsequently, the pulses is focused into the interior of a LiF crystal (size: 10×10×1 mm$^3$) through an objective lens (Mitutoyo, Plan Apo 50×, NA 0.55), with the focal point located approximately 0.1 mm beneath the sample surface, initiating localized photoexcitation. The other path is to send the pulse through a FACED module to generate a pulse train as the probe beam. This probe passes through a half-wave plate (HWP, LBTEK, AHWP20-SNIR) and polarizer (P1, Thorlabs, LPVIS100-MP2) to control its polarization direction, enabling a snapshot detection to the dynamic scene. It then enters the AOD-CUP system through another polarizer (P2, Thorlabs, LPVIS100-MP2) to acquire a 2D compressed image. In the experiment, the temporal spacing between adjacent sub-pulses in the probe sequence is approximately 400 ps, spanning a total observation window of ~4 ns. Each sub-pulse in the probe is linearly polarized, with P1 and P2 oriented orthogonally to enable visualization of transient stress wave propagation.

After data acquisition, the transient scene is reconstructed from the compressed measurement using the TV-CD algorithm. The reconstructed frames visualizing stress wave propagations captured by AOD-CUP are shown at the top panels of Figs. 5(b) and 5(c), which display the transient stress wave propagation along the (100) and (110) crystallographic orientations, respectively. For the (100) orientation, bright regions are observed propagating along the <110> direction, forming a distinct cross-shaped pattern surrounding the photoexcited region. This indicates the generation of strong stress waves along this crystallographic direction. Such a phenomenon is likely caused by rapid heating of the photoexcited region induced by nonlinear photoionization when the femtosecond laser is focused onto the LiF crystal[31]. The localized temperature increases sharply, and due to the constraint on thermal expansion, transient thermal stress is generated within the excitation volume. This stress is then rapidly released outward through an elastic relaxation process, launching stress waves that trigger lattice slip and the formation of dislocation bands[39]. Internal stress waves are observed approximately 1.2 ns after excitation and subsequently propagate outward. For comparison, the laser-induced dynamics for the (110) orientation differs from that of the (100) orientation where the brighter regions are observed to propagate along the <100> direction. Although internal stress waves are visible in the (100) orientation images, they are barely observed in the (110) orientation images. This suggests that the stress vector of the internal wave is oriented either parallel or perpendicular to the (110) plane.

To further verify the ultrafast imaging accuracy of the AOD-CUP system, we employed a narrowband bandpass filter (BPF, FWHM 2 nm) to filter the probe pulse train for a pump-probe imaging comparison. By adjusting the angle between the BPF plane and the pulse propagation direction, sub-pulses of different wavelengths (i.e., different time delays) were selectively transmitted. The corresponding stress wave images were then recorded using a conventional pump-probe microscope through multiple exposures, as shown at the bottom panel of Figs. 5(b) and (c). The results exhibit excellent agreement with those captured by AOD-CUP, demonstrating the accuracy and superior performance of the AOD-CUP system. Moreover, the propagation distances of primary stress waves was extracted from the reconstructed images along the <100> and <110> directions for the (100) and (110) orientation, as shown in Figs. 5(d) and 5(e). Meanwhile, we extracted the propagation distances of inner stress waves along the <100> and <110> directions from the (100) orientation, as shown in Fig. 5(f). By fitting the distance–time curves, the average propagation velocities in LiF along the <100> and <110> directions are determined to be 9.2 km/s and 8.7 km/s, respectively, while the inner stress wave propagates at an average speed of 5.1 km/s, approximately half that of the primary wave[40]. These results reveal the anisotropic nature of stress wave propagation in LiF, indicating direction-dependent velocities due to the variation of elastic properties along different crystallographic directions.

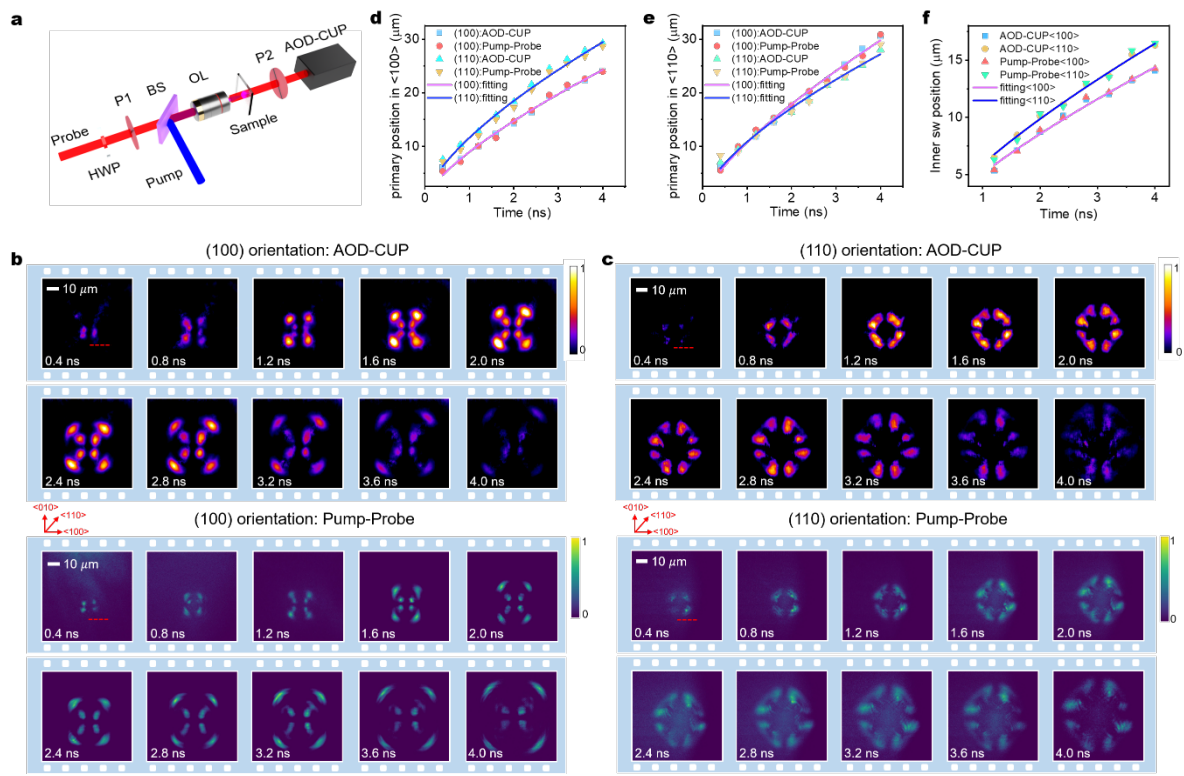

**Fig. 5.** AOD-CUP imaging of LiF crystals in different crystal directions excited by ultrafast laser pulses. (a) Experimental configuration; HWP: half-wave plate, P: polarizer, OL: objective lens, BS: beam splitter. (b) Reconstructed results of AOD-CUP in the (100) orientation compared with pump-probe results; (c) Same as (b) for (110) orientation; (d) Temporal evolution of the primary stress wave in the (100) and (110) orientation along the <100> direction by AOD-CUP and pump-probe results, together with fitting curve; (e) Same as (d) for <110> direction; (f) Temporal evolution of the inner stress wave in the (100) orientation along the <100> and <110> directions by AOD-CUP and pump-probe results, together with fitting curve; Scale bar: 10 μm.

### d. Observation of femtosecond laser-induced air plasma channels

To demonstrate the broad applicability of the AOD-CUP system in dynamic scenarios in various time scales, we further recorded the evolution of femtosecond laser-induced plasma channels in air at an increased imaging speed. As illustrated in Fig. 6(a), a 400 nm femtosecond laser pulse with a single-pulse energy of 500 μJ is focused into air by an objective lens (OL1, Mitutoyo, Plan Apo 10×). In this experiment, the temporal interval between adjacent sub-probe pulses was set to approximately 86 ps, resulting a total temporal window of ~860 ps. The laser-induced plasma evolution was imaged by another objective lens (OL2, Olympus, PLN 10×) and captured by the AOD-CUP system.

After data acquisition, the original scene was reconstructed using the TV-CD algorithm, as shown in Fig. 6(b). At 86 ps, the laser-induced plasma exhibits two filament-shaped structures. As time lapses, the plasma expands along the $y$-axis, ultimately evolving into a spindle-shaped structure with openings at both ends. This phenomenon is attributed to the high peak power of the excitation pulse, which drives the expansion of the plasma and displaces the surrounding air[41]. For comparison, we also employed conventional pump-probe microscopy as used in Fig. 5 to record the plasma channel evolution at different time delays. The morphology and dynamic evolution of the plasma channels captured by AOD-CUP show excellent agreement with those obtained by the pump-probe method. Furthermore, we extracted the plasma expansion distance along the $y$-direction (indicated by the red dashed line in the Fig. 6(b)) over time, as presented in Fig. 6(c), and calculated an average expansion velocity of approximately 31.5 km/s. The strong consistency between the results from the two techniques further validates the reliability of the AOD-CUP system.

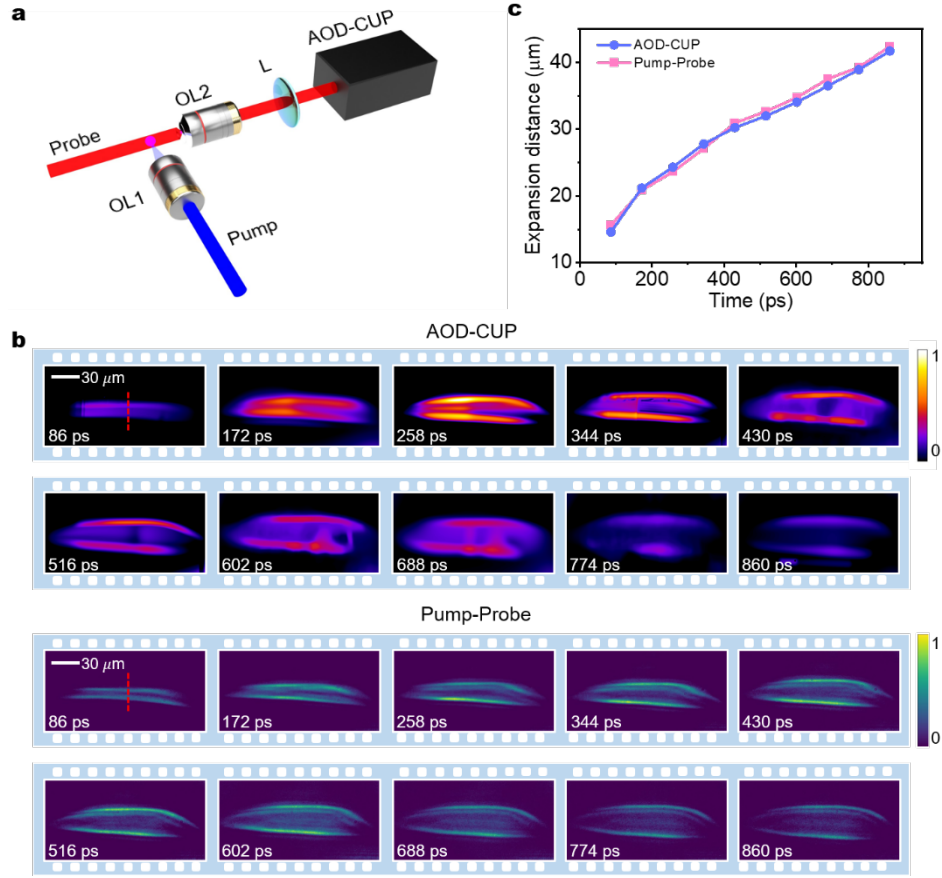

**Fig. 6.** AOD-CUP imaging of ultrafast laser pulse-induced air plasma. (a) Experimental configuration; OL: objective lens, L: lens. (b) Reconstructed results of AOD-CUP, compared with the results by pump-probe method; (c) Radial expansion distances of the plasma channel measured by AOD-CUP and pump-probe method. Scale bar: 30 μm.

### 4. Conclusion

In summary, we have developed an all-optical discrete-illumination compressed ultrafast photography technique, termed AOD-CUP. By employing FACED for ultrafast laser shaping, the input laser pulse is temporally stretched and shaped into a sequence of discrete pulses to illuminate transient scenes. This approach effectively modulates the data compression ratio, significantly suppresses interframe crosstalk, and circumvents the inherent charge effects of streak cameras, thereby enabling high-fidelity imaging of ultrafast phenomena in real time. Experimental results demonstrate that AOD-CUP offers superior spatial resolution and imaging quality. To demonstrate the superior imaging capabilities of AOD-CUP in capturing transient phenomena, we present its application in capturing the propagation of transient stress waves inside a LiF crystal induced by femtosecond laser irradiation, as well as the ultrafast evolution of plasma channels in air. These experiments visualize and elucidate the interaction mechanisms between femtosecond laser pulses and transparent solids, as well as the dynamic behaviors of laser-induced plasma. Moreover, AOD-CUP features a tunable interframe interval ranging from picoseconds to nanoseconds, showcasing its great potential in recording a wide variety of complex ultrafast dynamic events[42, 43].

Although AOD-CUP exhibits outstanding performance, there is still room for improvement in aspects such as imaging speed and dynamic range. Looking forward, the integration of deep-learning-based compressive sensing algorithms is expected to further enhance image quality. Additionally, coupling AOD-CUP with complementary techniques such as interferometry[44] may extend its capabilities to higher dimensional ultrafast imaging, offering powerful tools for advancing the frontiers of ultrafast science.

## Conflict of Interests

The authors declare no conflict of interest.

## Data Availability Statement

The data that support the findings of this study are available from the corresponding author upon reasonable request.